\documentclass[runningheads,openright]{svmult}

\usepackage{makeidx}   
\usepackage{graphicx}  
\usepackage{subeqnar}  
\usepackage{multicol}  
\usepackage{physprbb}  
\makeindex             

\newcommand{\age}{\mathrel{\hbox{\rlap{\hbox{\lower4pt\hbox{$\sim$}}}\hbox{$>$}}}}

\newcommand{\eg}{e.g., } 

\newcommand{\etal}{et al.}

\newcommand{\gae}{\mathrel{>\kern-1.0em\lower0.9ex \hbox{$\sim$}}}

\newcommand{\gsim}{\!\!\!\phantom{\ge}\smash{\buildrel{}\over {\lower2.5dd\hbox{$\buildrel{\lower2dd\hbox{$\displaystyle>$}}\over \sim$}}}\,\,}
\newcommand{\gtap}{\mathrel{\hbox{\rlap{\lower.55ex \hbox {$\sim$}}\kern-.3em \raise.4ex \hbox{$>$}}}}
\newcommand{\gtrsim}{\mathrel{\hbox{\rlap{\hbox{\lower4pt\hbox{$\sim$}}}\hbox{$>$}}}}

\newcommand{\kmsMpc}{\mbox{~km s$^{-1}$ Mpc$^{-1}$}}

\newcommand{\lae}{\mathrel{<\kern-1.0em\lower0.9ex \hbox{$\sim$}}}
\newcommand{\lesssim}{\mathrel{\hbox{\rlap{\hbox{\lower4pt\hbox{$\sim$}}}\hbox{$<$}}}}
\newcommand{\lsim}{\!\!\!\phantom{\le}\smash{\buildrel{}\over {\lower2.5dd\hbox{$\buildrel{\lower2dd\hbox{$\displaystyle<$}}\over \sim$}}}\,\,}

\newcommand{\ltap}{\mathrel{\hbox{\rlap{\lower.55ex \hbox {$\sim$}} \kern-.3em \raise.4ex \hbox{$<$}}}}
\newcommand{\ltsima}{$\; \buildrel < \over \sim \;$}

\newcommand{\simlt}{\lower.5ex\hbox{\ltsima}} 

\def\CFHT{{\sl Canada France Hawaii Telescope (CFHT)}\index{instruments, agencies, observatories, and programs!Canada France Hawaii Telescope (CFHT)}}

\def\CMB{{\sl Cosmic Microwave Background (CMB)}\index{Cosmic Microwave Background}}
\def\CMb{{\sl CMB}\index{Cosmic Microwave Background}}
\def\COBE{{\sl Cosmic Background Explorer (COBE)}\index{instruments, agencies, observatories, and programs!Cosmic Background Explorer (COBE)}}

\def\CTIO{{\sl Cerro Tololo Inter-American Observatory (CTIO)}\index{instruments, agencies, observatories, and programs!Cerro Tololo Inter-American Observatory (CTIO)}}
\def\CT{{\sl CTIO}\index{instruments, agencies, observatories, and programs!Cerro Tololo Inter-American Observatory (CTIO)}}
\def\CPA{{\sl Center for Particle Astrophysics}\index{instruments, agencies, observatories, and programs!Center for Particle Astrophysics}}
\def\CTSS{{\sl Calan/Tololo Supernova Search (CTSS)}\index{instruments, agencies, observatories, and programs!Calan/Tololo Supernova Search (CTSS)}}
\def\CTSs{{\sl CTSS}\index{instruments, agencies, observatories, and programs!Calan/Tololo Supernova Search (CTSS)}}

\def\E{{\sl Einstein}\index{instruments, agencies, observatories, and programs!Einstein Observatory -- HEAO-2 (Einstein)}}

\def\HZSNS{{\sl High-Z SN Search (HZSNS)}\index{instruments, agencies, observatories, and programs!High-Z SN Search (HZSNS)}}
\def\HZSNs{{\sl HZSNS}\index{instruments, agencies, observatories, and programs!High-Z SN Search (HZSNS)}}
\def\HST{{\sl Hubble Space Telescope (HST)}\index{instruments, agencies, observatories, and programs!Hubble Space Telescope (HST)}}
\def\H{{\sl HST}\index{instruments, agencies, observatories, and programs!Hubble Space Telescope (HST)}}

\def\I{{\sl IRAS}}

\def\KI{{\sl W.M. Keck Telescope I (KECK I)}\index{instruments, agencies, observatories, and programs!W.M. Keck Telescope)}}
\def\Ki{{\sl KECK I}\index{instruments, agencies, observatories, and programs!W.M. Keck Telescope}}

\def\Key{{Key Project}\index{instruments, agencies, observatories, and programs!Hubble Space Telescope (HST)!Key Project}}

\def\LBNL{{\sl Lawrence Berkeley National Laboratory (LBNL)}\index{instruments, agencies, observatories, and programs!Lawrence Berkeley National Laboratory (LBNL)}}
\def\LBNl{{\sl LBNL}\index{instruments, agencies, observatories, and programs!Lawrence Berkeley National Laboratory (LBNL)}}

\def\MAP{{\sl Microwave Anisotropy Probe (MAP)}\index{instruments, agencies, observatories, and programs!Microwave Anisotropy Probe (MAP)}}

\def\PLANCK{{\sl Planck}\index{instruments, agencies, observatories, and programs!PLANCK}}

\def\SCP{{\sl Supernova Cosmological Project (SCP)}\index{instruments, agencies, observatories, and programs!Supernova Cosmological Project (SCP)}}
\def\SCp{{\sl SCP}\index{instruments, agencies, observatories, and programs!Supernova Cosmological Project (SCP)}}

\def\SNAP{{\sl SuperNova/Acceleration Probe (SNAP)}\index{instruments, agencies, observatories, and programs!SuperNova/Acceleration Probe (SNAP)}}
\def\SNAp{{\sl SNAP}\index{instruments, agencies, observatories, and programs!SuperNova/Acceleration Probe (SNAP)}}

\def\SUBARU{{\sl Subaru Telescope)}\index{instruments, agencies, observatories, and programs!Subaru Telescope}}

\def\WFPC2{{\sl Wide-Field Planetary Camera 2 (WFPC2)}\index{instruments, agencies, observatories, and programs!Hubble Space Telescope (HST)!Wide-Field Planetary Camera 2 (WFPC2)}}
\def\WFPc2{{\sl WFPC2}\index{instruments, agencies, observatories, and programs!Hubble Space Telescope (HST)!Wide-Field Planetary Camera 2 (WFPC2)}}

\def\EPM{{expanding photosphere method (EPM)}\index{supernova!distance!expanding photosphere method}}
\def\EPm{{EPM}\index{supernova!distance!expanding photosphere method}}

\def\MLCS{{\sl multi-color light curve shape method (MLCS)}\index{supernova!light curve!multi-color light curve shape method (MLCS)}}
\def\MLCs{{\sl MLCS}\index{supernova!light curve!multi-color light curve shape method (MLCS)}}

\def \aap #1 #2 {{Astron. Astrophys.\/} {\bf #1}, #2~}
\def \aar #1 #2 {{Astron. Astrophys. Rev.\/} {\bf #1}, #2~}
\def \aas #1 #2 {{Astron. Astrophys. Suppl. Ser.\/} {\bf #1}, #2~}
\def \aj #1 #2 {{Astron. J.\/} {\bf #1}, #2~}
\def \al #1 #2 {{Astron. Lett.\/} {\bf #1}, #2~}
\def \an #1 #2 {{Astron. Nach.\/} {\bf #1}, #2~}
\def \annap #1 #2 {{Annals Ap.\/} {\bf #1}, #2~}
\def \aph #1 {{astro-ph\/} {#1}~}
\def \ar #1 #2 {{Astron. Rep.\/} {\bf #1}, #2~}
\def \araap #1 #2 {{Ann. Rev. Astron. Astrophys.\/} {\bf #1}, #2~}
\def \asiagoc #1 #2 {{Asiago Contr.\/} {\bf #1}, #2~}
\def \apj #1 #2 {{Astrophys. J.\/} {\bf #1}, #2~}
\def \apjl #1 #2 {{Astrophys. J. Lett.\/} {\bf #1}, #2~}
\def \apjs #1 #2 {{Astrophys. J. Suppl.\/} {\bf #1}, #2~}
\def \apjsub #1 {{Astrophys. J.\/} {#1}~}
\def \apph #1 #2 {{Astropart. Phys.\/} {\bf #1}, #2~}
\def \apss #1 #2 {{Astrophys. Space Sci.\/} {\bf #1}, #2~}
\def \aspc #1 #2 {{ASP Conf.~Proc.\/} {\bf #1}, #2~}
\def \aspl #1 {{ASP Leaflet\/} {#1}~}
\def \asr #1 #2 {{Adv. Space Res.\/} {\bf #1}, #2~}
\def \astrl #1 #2 {{Astron. Lett.\/} {\bf #1}, #2~}
\def \azh #1 #2 {{Astron. Zhurnal\/} {\bf #1}, #2~}
\def \baas #1 #2 {{Bull. Am. Astron. Soc.\/} {\bf #1}, #2~}
\def \ban #1 #2 {{Bull. Astron. Inst. Neth.\/} {\bf #1}, #2~}
\def \basi #1 #2 {{Bull. Astron. Soc. India\/} {\bf #1}, #2~}
\def \ca #1 #2 {{Chinese Astron.\/} {\bf #1}, #2~}
\def \coap #1 #2 {{Contrib. Oss. Astrofis. Padova in Asiago\/} {\bf #1}, #2~}
\def \cap #1 #2 {{Comm. Astrophys.\/} {\bf #1}, #2~}
\def \emsg #1 {{ESO Messenger\/} {#1}~}
\def \gcn #1 {{GCN\/} {#1}~}
\def \hast #1 #2 {{Highlights of Astronomy\/} {\bf #1}, #2~}
\def \iauc #1 {{IAUC\/} {#1}~}
\def \iaus #1 #2 {{IAU Symp. 110: VLBI \& Compact Radio Sources\/} {\bf #1}, #2~}
\def \inprep {{in preparation\/}~}
\def \jcam #1 #2 {{J. Comp. Appl. Math.\/} {\bf #1}, #2~}
\def \jet #1 #2 {{JETP Lett.\/} {\bf #1}, #2~}
\def \jha #1 #2 {{J. Hist. Astron.\/} {\bf #1}, #2~}
\def \jrasc #1 #2 {{J. R. Astron. Soc. Canada\/} {\bf #1}, #2~}
\def \mem #1 #2 {{Mem. R. Astron. Soc.\/} {\bf #1}, #2~}
\def \mess #1 #2 {{The Messenger\/} {\bf #1}, #2~}
\def \mnras #1 #2 {{Mon. Not. R. Astron. Soc.\/} {\bf #1}, #2~}
\def \mplb #1 #2 {{Mod. Phys. Lett. B\/} {\bf #1}, #2~}
\def \nat #1 #2 {{Nature\/} {\bf #1}, #2~}
\def \newa #1 #2 {{New Astron.\/} {\bf #1}, #2~}
\def \nuca #1 #2 {{Nucl. Phys. A\/} {\bf #1}, #2~}
\def \nucb #1 #2 {{Nucl. Phys. B\/} {\bf #1}, #2~}
\def \npps #1 #2 {{Nucl. Phys. Proc. Suppl.\/} {\bf #1}, #2~}
\def \nyasa #1 #2 {{NY Acad. Sci. Ann.\/} {\bf #1}, #2~}
\def \obsy #1 #2 {{The Observatory\/} {\bf #1}, #2~}
\def \phfl #1 #2 {{Phys. Fluids\/} {\bf #1}, #2~}
\def \phytd #1 #2 {{Phys. Today\/} {\bf #1}, #2~}
\def \prl #1 #2 {{Phys. Rev. Lett.\/} {\bf #1}, #2~}
\def \prp #1 #2 {{Phys. Rep.\/} {\bf #1}, #2~}
\def \phyr #1 #2 {{Phys. Rev.\/} {\bf #1}, #2~}
\def \phyrd #1 #2 {{Phys. Rev. D\/} {\bf #1}, #2~}
\def \prasa #1 #2 {{Proc. Astron. Soc. Australia\/} {\bf #1}, #2~}
\def \pasa #1 #2 {{Pub. Astron. Soc. Australia\/} {\bf #1}, #2~}
\def \pasj #1 #2 {{Pub. Astron. Soc. Japan\/} {\bf #1}, #2~}
\def \pasp #1 #2 {{Pub. Astron. Soc. Pacific\/} {\bf #1}, #2~}
\def \qjras #1 #2 {{Q. J. R. Astron. Soc.\/} {\bf #1}, #2~}
\def \rma #1 #2 {{Rev. Mod. Astron.\/} {\bf #1}, #2~}
\def \rpp #1 #2 {{Rep. Prog. Phys.\/} {\bf #1}, #2~}
\def \rpph #1 #2 {{Rev. Plasma Phys.\/} {\bf #1}, #2~}
\def \sait #1 #2 {{Mem.\ Soc.\ Astron.\ It.\/} {\bf #1}, #2~}
\def \sast #1 #2 {{Sov. Astron.\/} {\bf #1}, #2~}
\def \sal #1 #2 {{Sov. Astron. Lett.\/} {\bf #1}, #2~}
\def \sat #1 #2 {{Sky \& Tel.\/} {\bf #1}, #2~}
\def \sci #1 #2 {{Science\/} {\bf #1}, #2~}
\def \spie #1 #2 {{SPIE\/} {\bf #1}, #2~}
\def \shns #1 #2 {{Stud. Hist. Nat. Sci.\/} {\bf #1}, #2~}
\def \va #1 #2 {{Vist. Astron.\/} {\bf #1}, #2~}
\newcommand{\GAL}[1]{#1\index{galaxy!individual!#1}}

\newcommand{\SN}[1]{SN#1\index{supernova!individual!SN#1}}

\newcommand{\SNt}[1]{SN#1\index{supernova!type!SN#1}}
\newcommand{\type}[1]{type #1\index{supernova!type!SN#1}}
\index{agency|see {instruments, agencies, observatories, and programs}}
\index{Baade-Wesselink method|see {supernova, distance, Baade-Wesselink method}}
\index{blastwave|see {(supernova, shock, circumstellar) or (GRB, shock, external)}}
\index{circumstellar shock|see {supernova, shock, circumstellar}}
\index{CLUMPYSYN|see {supernova, model, CLUMPYSYN}}
\index{collapsar|see {GRB, progenitor, collapsar}}
\index{dark burst|see {GRB, optical, dark}}
\index{expanding photosphere method|see {supernova, distance, expanding photosphere method}}
\index{filled-center supernova remnant|see {supernova remnant, plerionic}}
\index{forward shock|see {(supernova, shock, circumstellar) or (GRB, shock, external)}}
\index{GRB!model!interstellar medium interactor|see {GRB, circumburst medium, interstellar medium}}
\index{GRB!model!wind interactor|see {GRB, circumburst medium, stellar wind}}
\index{guest star|see {supernova, historical}}
\index{GRB!progenitor!hypernova|see {hypernova}}
\index{GRB!progenitor!Wolf-Rayet star|see {star, Wolf-Rayet}}
\index{interstellar scintillation|see {GRB, radio, scintillation}}
\index{ML MONTE CARLO|see {supernova, model, ML MONTE CARLO}}
\index{NLTE (non-local thermodynamic equilibrium)|see {supernova, model, NLTE}}
\index{outgoing shock|see {(supernova, shock, circumstellar) or (GRB, shock, external)}}
\index{observatory|see {instruments, agencies, observatories, and programs}}
\index{PHOENIX|see {supernova, model, PHOENIX}}
\index{presupernova wind|see {supernova, progenitor, wind}}
\index{program|see {instruments, agencies, observatories, and programs}}
\index{pulsar|see {star, neutron}}
\index{radio supernova|see {supernova, radio}}
\index{RL NEBULAR|see {supernova, model, RL NEBULAR}}
\index{Sobolev approximation|see {supernova, model, Sobolev approximation}}
\index{supernova!cosmology|see {cosmology}}
\index{synchrotron self-absorption|see {supernova, synchrotron, self-absorption}}
\index{SYNOW|see {supernova, model, SYNOW}}
\index{synthetic spectrum|see {supernova, model, synthetic spectra}}
\index{VLBI|see {instruments, agencies, observatories, and programs, VLBI}}

\begin{document}


\mainmatter

\title*{Measuring cosmology\index{cosmology} with Supernovae}
\toctitle{Measuring cosmology with Supernovae}
%
%
\titlerunning{Measuring cosmology\index{cosmology} with Supernovae}
%
\author{Saul Perlmutter\inst{1}
\and Brian P. Schmidt\inst{2}}
\authorrunning{Perlmutter and Schmidt}
%
%
\institute{Physics Division, Lawrence Berkeley National Laboratory,
University of California,
Berkeley, CA 94720, USA 
\and Research School of Astronomy and Astrophysics, The Australian National University, via Cotter Rd, Weston Creek, ACT 2611, Australia}

\maketitle              

\begin{abstract}
Over the past decade, supernovae have emerged as some of the most powerful tools for measuring extragalactic distances.  A well developed physical understanding of \type{II} supernovae allow them to be used to measure distances independent of the extragalactic distance scale. Type Ia supernovae are empirical tools whose precision and intrinsic brightness make them sensitive probes of the cosmological expansion. Both types of supernovae are consistent with a Hubble Constant within $\sim$10\% of $H_0 = 70$ \kmsMpc.  Two teams have used \type{Ia} supernovae to trace the expansion of the Universe to a look-back time more than 60\% of the age of the Universe. These observations show an accelerating Universe which is currently best explained by a cosmological constant or other form of dark energy with an equation of state near $w = p/\rho = -1$. While there are many possible remaining systematic effects, none appears large enough to challenge these current results. Future experiments are planned to better characterize the equation of state of the dark energy leading to the observed acceleration by observing hundreds or even thousands of objects. These experiments will need to carefully control systematic errors to ensure future conclusions  are not dominated by effects unrelated to cosmology.
\end{abstract}

\section{Introduction}

Understanding the global history of the Universe is a fundamental goal of cosmology\index{cosmology}. One of the conceptually simplest tests in the repertoire of the cosmologist is observing how a standard candle\index{standard candle} dims as a function of redshift. The nearby Universe provides the current rate of expansion, and with more distant objects it is possible to start seeing the varied effects of cosmic curvature and the Universe's expansion history (usually expressed as the rate of acceleration/deceleration).   Over the past several decades a paradigm for understanding the global properties of the Universe has emerged based on General Relativity\index{General Relativity} with the assumption of a homogeneous and isotropic Universe.  The relevant constants in this model are the Hubble constant\index{Hubble constant} (or current rate of cosmic expansion),  the relative fractions of species of matter that contribute to the energy density of the Universe, and these species' equation of state. 

Early luminosity distance investigations used the brightest objects available for measuring distance -- bright galaxies \cite{B57,HMS56}, but these efforts were hampered by the impreciseness of the distance indicators and the changing properties of the distance indicators as a function of look back time. Although many other methods for measuring the global curvature\index{cosmology!Universe!topology} and cosmic deceleration\index{cosmology!Universe!deceleration} exist (see, \eg \cite{Peebles93}), supernovae (SNe) have emerged as one of the preeminent distance methods due to their significant intrinsic brightness (which allows them to be observable in the distant Universe), ubiquity (they are visible in both the nearby and distant Universe), and their precision (\type{Ia} SNe provide distances that have a precision of approximately 8\%). 

\section{Supernovae as Distance Indicators}

\subsection{Type II Supernovae and the Expanding Photosphere Method}

Massive stars\index{star!high mass} come in a wide variety of luminosities and sizes and would seemingly not be useful objects for making distance\index{supernova!distance!expanding photosphere method} measurements under the standard candle\index{standard candle} assumption.  However, from a radiative transfer standpoint these objects are relatively simple and can be modeled with sufficient accuracy to measure distances to approximately 10\%.  The \EPM, was developed by Kirshner and Kwan \cite{KK74}, and implemented on a large number of objects by Schmidt \etal\ \cite{S94} after considerable improvement in the theoretical understanding of \type{II} SN (\SNt{II}) atmospheres \cite{EK89,ESK96,wagoner81}. 

\EPm\ assumes that \SNt{II} radiate as dilute blackbodies

\begin{equation}
\theta_{ph} = {R_{ph}\over D} = \sqrt{ {F_\lambda \over {\zeta^2 \pi B_\lambda(T)}}},\label{eq:epm1}
\end{equation}

\noindent where $\theta_{ph}$ is the angular size of the photosphere of the SN, $R_{ph}$ is the radius
of the photosphere, $D$ is the distance to the SN, $F_{\lambda}$ is the observed flux density of the SN, and
$B_\lambda(T)$ is the Planck function\index{Planck function} at a temperature $T$. Since \SNt{II} are not perfect blackbodies, we
include a correction factor, $\zeta$, which is calculated from radiate transfer models of \SNt{II}. SNe freely
expand, and

\begin{equation}
R_{ph} = v_{ph}(t-t_0)+R_0, \label{eq:epm2}
\end{equation}

\noindent where $v_{ph}$ is the observed velocity of material at the position of the photosphere, and $t$ is the time elapsed
since the time of explosion, $t_0$. For most stars, the stellar radius ,$R_0$, at the time of explosion is negligible,
and Eqs.~(\ref{eq:epm1}--\ref{eq:epm2}) can be combined to yield

\begin{equation}
 t = D\left( {\theta_{ph} \over v_{ph}} \right ) + t_0 \label{eq:epm}
\end{equation}

By observing a \SNt{II} at several epochs,  measuring the flux density and temperature of the SN (via broad band
photometry) and $v_{ph}$ from the minima of the weakest lines in the SN spectrum, we can solve simultaneously for
the time of explosion and distance to the \SNt{II}. The key to successfully measuring distances\index{supernova!distance} via \EPm\ is an
accurate calculation of $\zeta (T)$. Requisite calculations were performed by Eastman \etal\ \cite{ESK96} but, unfortunately, no other calculations of $\zeta(T)$ have yet been published for typical \SNt{IIP} progenitors.
 
Hamuy \etal\ \cite{h2001} and Leonard \etal\ \cite{leonard2002} have measured the distances to
\SN{1999em}, and they have investigated other aspects of \EPm. Hamuy \etal\ \cite{h2001} challenged the
prescription of measuring velocities from the minima of weak lines and developed a framework of cross correlating
spectra with synthesized spectra to estimate the velocity of material at the photosphere. This different
prescription does lead to small systematic differences in estimated velocity using weak lines but,
provided the modeled spectra are good representations of real objects, this method should be more correct.
At present, a revision of the \EPm\ distance scale using this method of estimating $v_{ph}$ has
not been made. 

Leonard \etal\ \cite{leonard2001} have obtained spectropolarimetry of \SN{1999em} at many epochs
and see polarization intrinsic to the SN which is consistent with the SN have asymmetries of $10-20$\%.
Asymmetries at this level  are found in most \SNt{II} \cite{Wang2001}, and may ultimately limit the 
accuracy \EPm\ can achieve on a single object ($\sigma \sim 10\%$).  However, the mean of all \SNt{II} distances should remain 
unbiased.

Type II SNe have played an important role in measuring the Hubble constant\index{Hubble constant} independent of the rest of the
extragalactic distance scale. In the next decade it is quite likely that surveys will begin to turn up significant
numbers of these objects at $z \sim 0.5$ and, therefore, the possibility exists that \SNt{II} will be able to make 
a contribution to the measurement of cosmological parameters beyond the Hubble Constant. Since \SNt{II} do not have the
precision of the \SNt{Ia} (next section) and are significantly harder to measure, they will not replace
the \SNt{Ia} but will remain an independent class of objects which have the potential to confirm the interesting results that
have emerged from the \SNt{Ia} studies.

\subsection{Type Ia Supernovae as Standardized Candles}

\SNt{Ia} have been used as extragalactic distance indicators since Kowal 
\cite{K68} first published his Hubble diagram\index{Hubble diagram} ($\sigma = 0.6$ mag) for 
\type{I} SNe. We now recognize that the old \type{I} SNe spectroscopic class is 
comprised of two distinct physical entities: \SNt{Ib/c} which are massive 
stars\index{star!high mass} that undergo core collapse (or in some rare cases might undergo a 
thermonuclear detonation\index{supernova!thermonuclear} in their cores) after losing their hydrogen 
atmospheres, and \SNt{Ia} which are most likely thermonuclear 
explosions of white dwarfs\index{star!white dwarf}.  In the mid-1980s it was recognized that 
studies of the \type{I} SN sample had been confused by these 
similar appearing SNe, which were henceforth classified as type 
Ib \cite{panagia85,uomoto85,wheellev85} 
and \type{Ic} \cite{harkness90}.  By the late 
1980s/early 1990s, a strong case was being made 
that the vast majority of the true \type{Ia} 
SNe had strikingly similar light curve\index{supernova!light curve} shapes \cite{cadonau87,leib88,leibtammann90,leibetal91}, spectral time series \cite{branch89,filippenko91,hamuy91,pearce88b}, and absolute 
magnitudes \cite{leibtammann90,millerbranc90}. 
There were a small minority of clearly peculiar \type{Ia} SNe (\eg \SN{1986G} \cite{P87}, \SN{1991bg} \cite{Fetal92,Letal93}, and \SN{1991T} 
\cite{Fetal92,Petal92}), but 
these could be identified and removed by their unusual spectral 
features.  A 1992 review by Branch and Tammann 
\cite{branchtammann92} of a variety of 
studies in the literature concluded that the intrinsic dispersion in 
B and V maximum for \type{Ia} SNe must be $<0.25$ mag, 
making them ``the best standard candles\index{standard candle} known so far.''

In fact, the Branch and Tammann review indicated that the magnitude 
dispersion was probably even smaller, but the measurement uncertainties 
in the available datasets were too large to tell.   The 
\CTSS, a program begun by Hamuy \etal\ 
\cite{H93} in 1990, took the field a dramatic step forward by 
obtaining a crucial set of high quality SN light curves and 
spectra.  By targeting a magnitude range that would discover \type{Ia} 
SNe in the redshift range $z = 0.01-0.1$, the \CTSs\ 
was able to compare the peak magnitudes of SNe whose 
relative distance could be deduced from their Hubble velocities.

The \CTSs\ observed some 25 fields (out of a 
total sample of 45 fields) twice a month for over three and one half years 
with photographic plates or film at the \CTIO\ Curtis Schmidt
telescope, and then organized extensive follow-up photometry campaigns 
primarily on the \CT\ 0.9 m telescope, and spectroscopic observation on 
either the \CT\ 4 m or 1.5 m telescope.  Toward the end of this search, Hamuy 
\etal\ \cite{H93} pointed out the difficulty of this comprehensive 
project:  ``Unfortunately, the appearance of a SN is not predictable.  
As a consequence of this we cannot schedule the followup observations 
{\em a priori}, and we generally have to rely on someone else's 
telescope time.   This makes the execution of this project somewhat 
difficult.''   Despite these challenges, the search was a major success;  
with the cooperation of many visiting \CT\ astronomers and \CT\ staff, 
it contributed 30 new \type{Ia} SN light curves to the pool 
\cite{H95} with an almost unprecedented control of measurement 
uncertainties.

As the \CTSs\ data began to become available, several methods 
were presented that could select for the ``most standard'' subset of 
the \type{Ia} standard candles, a subset which remained the dominant 
majority of the ever-growing sample \cite{branchfisher93}.   For example, 
Vaughan \etal\ \cite{vaughn95} presented a cut on the B-V color at 
maximum that would select what were later called the ``Branch Normal'' 
\SNt{Ia}, with an observed dispersion of less than 0.25 mag.   

Phillips \cite{P93} found a tight correlation between the rate at which 
a \type{Ia} SN's luminosity declines and its absolute magnitude, 
a relation which apparently applied not only to the Branch Normal\index{supernova!light curve!Branch Normal} \type{Ia} SNe, but also to the peculiar \type{Ia} SNe.  Phillips 
plotted the absolute magnitude of the existing set of nearby \SNt{Ia}, 
which had dense photoelectric or  CCD coverage, versus the parameter 
$\Delta m_{15} ({\rm B})$, the amount the SN decreased in brightness in the 
B-band  over the 15 days following maximum light. The sample showed a 
strong correlation which, if removed, dramatically improved the 
predictive power of \SNt{Ia}.  Hamuy \etal\ \cite{H96c} used this empirical 
relation to reduce the scatter in the Hubble diagram\index{Hubble diagram} to $\sigma < 
0.2$~mag in V for a sample of nearly 30 \SNt{Ia} from the \CTSs\ search.

Impressed by the success of  the $\Delta m_{15}({\rm B})$ parameter, Riess \etal\ \cite{RPK96} 
developed the \MLCS, which 
parameterized the shape of SN light curves as a function of their absolute 
magnitude at maximum. This method also included a sophisticated error model 
and fitted observations in all colors simultaneously, allowing a color excess 
to be included. This color excess, which we attribute to intervening dust, 
enabled the extinction to be measured.  Another method that has been used widely
in cosmological measurements with \SNt{Ia} is the ``stretch'' method\index{supernova!light curve!stretch method} described
in Perlmutter \etal\ \cite{Perl97,Perl99}.  This method is 
based on the observation that the entire range of \SNt{Ia} light curves, at least in the B 
and V-bands, can be represented with a simple time stretching (or shrinking) of a 
canonical light curve. The coupled stretched B and V light curves serve
as a parameterized set of light curve shapes \cite{goldhaber01}, providing many of the benefits of the
MLCS method but as a much simpler (and constrained) set.  This method, as well as recent implementations
of $\Delta m_{15} ({\rm B})$ \cite{Germany02,Phillips99}, also allows extinction to be directly
incorporated into the \SNt{Ia} distance\index{supernova!distance} measurements. 
Other methods that correct for intrinsic luminosity differences or limit the input sample by various
criteria have also been proposed to increase the precision of \type{Ia} SNe as distance indicators \cite{B96,F95,TS95,v95}.  While these latter techniques are not as developed as  the $\Delta m_{15} ({\rm B})$,
\MLCs, and stretch methods, they all provide distances that are comparable
in precision, roughly $\sigma = 0.18$ mag about the inverse square law, equating to a fundamental precision of
\SNt{Ia} distances of $\sim6$\% (0.12 mag), once photometric uncertainties and peculiar velocities are removed.
Finally, a ``poor man's'' distance indicator, the snapshot method\index{supernova!light curve!snapshot method} \cite{Riesssnapshot}, combines information contained
in one or more SN spectra with as little as one night's multi-color photometry. This method's accuracy depends 
critically on how much information is available.

\section{Cosmological Parameters}

The standard model for describing the global evolution of the Universe is based 
on two equations that make some simple, and hopefully valid, assumptions.  If 
the Universe is isotropic and homogenous on large scales,  the 
Robertson-Walker Metric\index{Robertson-Walker Metric},

\begin{equation}
ds^2 = dt^2-a(t)\left[{dr^2\over{1-kr^2}}+r^2d\theta^2\right].\label{eq:RW}
\end{equation}

\noindent gives the line element distance(s) between two objects with coordinates $r$,$\theta$ 
and time separation, $t$. The Universe\index{cosmology!Universe!topology} is assumed to have a simple topology such 
that, if it has negative, zero, or positive curvature, $k$ takes the value 
${-1,0,1}$, respectively. These models of the Universe are said to be open, flat, 
or closed, respectively.  The dynamic evolution of the Universe needs to be
input into the Robertson-Walker Metric by the specification of the scale factor $a(t)$, 
which gives the radius of curvature of the Universe over time -- or more simply, 
provides the relative size of a piece of space at any time. This description of 
the dynamics of the Universe is derived from General Relativity\index{General Relativity}, and is known as the
Friedman equation\index{Friedman equation}

\begin{equation}
H^2\equiv(\dot a/a)^2 = {8\pi G\rho\over 3} - {k\over a^2}. \label{eq:F}
\end{equation}

The expansion rate\index{cosmology!Universe!expansion rate} of our Universe ($H$), is called the Hubble parameter\index{Hubble parameter} (or the 
Hubble constant\index{Hubble constant}, $H_0$, at the present epoch) and depends on the  
content of the Universe.  Here we assume the Universe is composed of a set of 
components, each having a fraction, $\Omega_i$, of the critical density\index{cosmology!Universe!critical density} 

\begin{equation}
\Omega_i = {{\rho_i}\over{\rho_{crit}}} = {{\rho_i}\over{ {3 H_0^2}\over{8\pi G}}},\label{eq:Om}
\end{equation}

\noindent with an equation of state which relates the density, $\rho_i$, and pressure, $p_i$, as
$w_i = p_i/\rho_i$. For example, $w_i$ takes the value 0 for normal matter, +1/3 for photons, and -1 for 
the cosmological constant\index{cosmology!cosmological constant}. The equation of state parameter does not need to 
remain fixed; if scalar fields are present, the effective $w$ will change over time.
Most reasonable forms of matter or scalar fields  have $w_i \ge -1$, although nothing seems manifestly
forbidden. Combining Eqs.~(\ref{eq:RW}--\ref{eq:Om}) yields solutions to the global evolution of the Universe \cite{CL}. 

The luminosity distance\index{cosmology!luminosity distance}, $D_L$, which is defined as the apparent brightness of
an object as a function of its redshift $z$ -- the amount an object's light has
been stretched by the expansion of the Universe -- can be derived
from Eqs.~(\ref{eq:RW}--\ref{eq:Om}) by solving for the surface area as
a function of $z$, and taking into account the effects of time dilation\index{cosmology!time dilation} 
\cite{goldhaber97,goldhaber01,Letal96,Riess97} and energy dimunition 
as photons get stretched traveling through the expanding Universe.  $D_L$ is given by the numerically integrable equation,  

\begin{equation}
 D_L = {c\over H_0} (1+z)\kappa_0^{-1/2}S\lbrace\kappa_0^{1/2}
\int_0^zdz'[\sum_i
\Omega_i(1+z')^{3+3w_i}-\kappa_0(1+z')^2 ]^{-1/2}\rbrace. \label{eq:DL}
\end{equation}

\noindent $S(x) = \sin(x)$, $x$, or $\sinh(x)$ for closed, flat\index{cosmology!Universe!curvature}, and 
open models respectively, and the curvature parameter $\kappa_0$, is defined as 
$\kappa_0 = \sum_i\Omega_i-1$.
                      
Historically, Eq.~(\ref{eq:DL}) has not been easily integrated and has been expanded in a Taylor series to give 

\begin{equation}
 D_L = {c \over H_0}\lbrace z+z^2\left({1-q_0\over 2}\right)  +{\cal O}(z^{3})\rbrace, \label{eq:expansion}
\end{equation}

\noindent where the deceleration parameter\index{cosmology!Universe!deceleration}, $q_0$, is given by

\begin{equation}
 q_0 = {1\over 2}\sum_i\Omega_i(1+3w_i).\label{eq:q}
\end{equation}

From Eq.~(\ref{eq:q}) we can see that, in the nearby Universe, the
luminosity distances\index{cosmology!luminosity distance} scale linearly with redshift, with $H_0$ serving as
the constant of proportionality. In the more distant Universe, $D_L$ depends 
to first order on the rate of acceleration/deceleration ($q_0$) or, 
equivalently, on the amount and types of matter that make up the Universe.
For example, since normal matter has $w_M = 0$ and the cosmological constant\index{cosmology!cosmological constant}
has $w_\Lambda = -1$, a universe composed of only these two forms of 
matter/energy has $q_0 = \Omega_M/2 - \Omega_\Lambda$. In a universe composed of
these two types of matter, if $\Omega_\Lambda < \Omega_M/2$, $q_0$ is positive
and the universe is decelerating\index{cosmology!Universe!deceleration}. These decelerating universes have $D_L$ smaller as a function of $z$ than their accelerating counterparts\index{cosmology!Universe!acceleration}.

If distance measurements are made at a low-$z$ and a small range of redshift at higher $z$ (\eg $0.3 > z > 0.5$),
there is a degeneracy between $\Omega_M$ and $\Omega_\Lambda$.  It is impossible to pin down
the absolute amount of either species of matter.  One can only determine their relative dominance, which, at $z = 0$, is given
by Eq.~(\ref{eq:q}). However, Goobar and Perlmutter \cite{GoobPerl95} pointed out that 
by observing objects over a larger range of high redshift (\eg $0.3 > z > 1.0$) this 
degeneracy can be broken, providing a measurement of the absolute fractions of
$\Omega_M$ and $\Omega_\Lambda$.

\begin{figure}[t]
\centering
\centering \includegraphics[width=10cm]{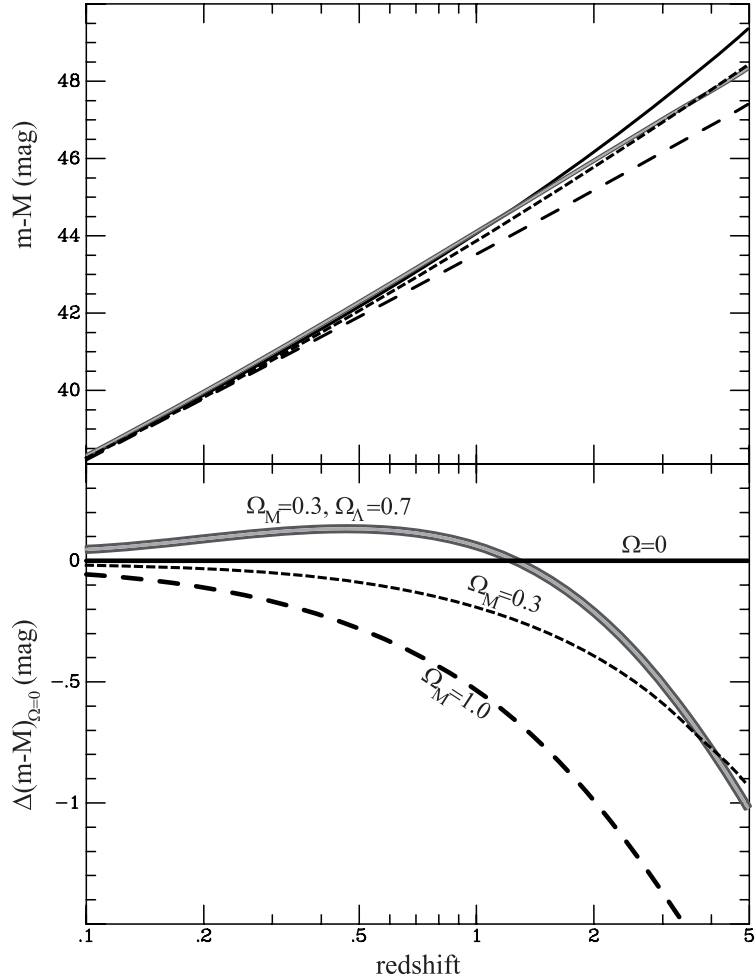}
\caption{\label{fig:DLvsz} $D_L$ expressed as distance modulus $(m-M)$ for four relevant cosmological models; $\Omega_M = 0$, $\Omega_\Lambda = 0$ (empty Universe, {\it solid line}); $\Omega_M = 0.3$, $\Omega_\Lambda = 0$ ({\it short dashed line}); $\Omega_M = 0.3$, $\Omega_\Lambda = 0.7$ ({\it hatched line}); and $\Omega_M = 1.0$, $\Omega_\Lambda = 0$ ({\it long dashed line}).  In the bottom panel the empty Universe has been subtracted from the other models to highlight the differences.}
\end{figure}

\begin{figure}[t]
\centering
\centering \includegraphics[width=10cm]{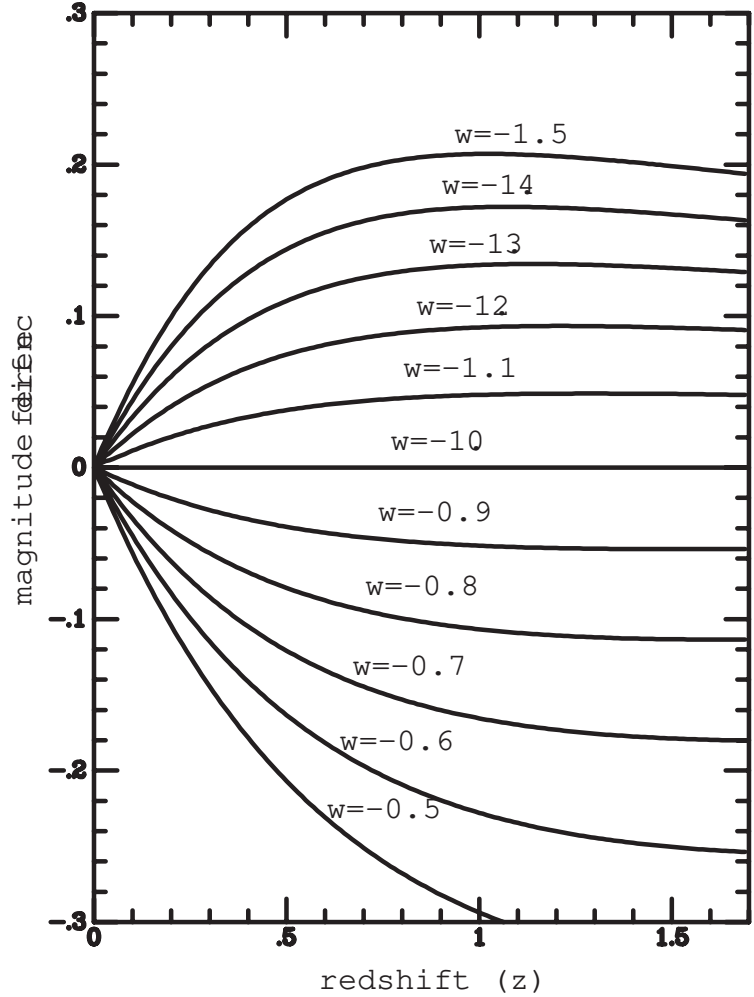}
\caption{\label{fig:alpha} $D_L$ for a variety of cosmological models containing 
$\Omega_M = 0.3$ and $\Omega_x = 0.7$ with a constant (not time-varying) equation of state $w_x$. The $w_x = -1$ model has been
subtracted off to highlight the differences between the various models}
\end{figure} 

To illustrate the effect of cosmological parameters on the luminosity distance, 
in Fig.~\ref{fig:DLvsz} we plot a series of models for both $\Lambda$ and non-$\Lambda$ universes.
In the top panel, the various models show the same linear behavior at $z < 0.1$ with
models having the same $H_0$  being indistinguishable to a few percent. By 
$z = 0.5$ the models with significant $\Lambda$ are clearly separated, with luminosity distances that 
are significantly further than the zero-$\Lambda$ universes.  Unfortunately, two 
perfectly reasonable universes, given our knowledge of the local matter density 
of the Universe ($\Omega_M \sim 0.2$), one with a large cosmological constant\index{cosmology!cosmological constant},
$\Omega_\Lambda = 0.7$, $\Omega_M = 0.3$ and one with no cosmological constant, $\Omega_M = 0.2$,
show differences of less than 25\%, even to redshifts of $z > 5$. Interestingly, the 
maximum difference between the two models is at $z \sim 0.8$, not at large $z$.   Fig.~\ref{fig:alpha} illustrates the effect of changing the equation of state of 
the non-matter, dark energy\index{cosmology!dark energy} component, assuming a flat universe\index{cosmology!Universe!topology}, 
$\Omega_{tot} = 1$.  If we are to discern a dark energy component that is not a 
cosmological constant, measurements better than 5\% are clearly required, 
especially since the differences in this diagram include the assumption of 
flatness and also fix the value of $\Omega_M$.  In fact, to discriminate 
among the full range of dark energy models with time varying equations of state
will require much better accuracy than even this challenging goal.

\section{Measuring the Hubble Constant}

Schmidt \etal\ \cite{S94}, using a sample of 16 \SNt{II}, 
estimated $H_0 = 73 \pm 6$(statistical)$\pm 7$ (systematic) using \EPm.
This estimate is independent of other rungs in the extragalactic
distance ladder, the most important of which are the Cepheids\index{star!Cepheid}, which currently
calibrate most other distance methods (such as \SNt{Ia}). 
The Cepheid and \EPm\ distance scales, compared galaxy to galaxy, agree to within 5\% and
are consistent within the errors \cite{ESK96,leonard2002}.  This provides confidence that both methods are providing accurate distances.

\begin{figure}[t]
\centering \includegraphics[width=12cm]{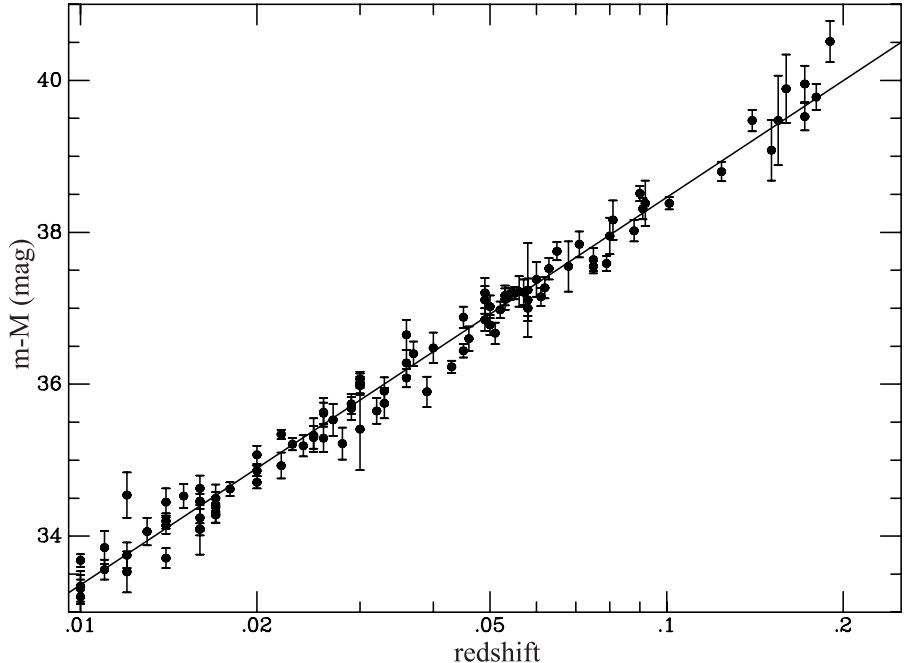}
\caption{The Hubble diagram for \SNt{Ia} from $0.01 > z > 0.2$ \cite{Germany02,H96c,Jha02,Riess99a}. The 102 objects in this range have a residual about the inverse
square line of $\sim10$\%.}
\label{fig:h0}
\end{figure}

The current nearby \SNt{Ia} sample \cite{Germany02,H95,Jha02,Riess99a} contains more than 100 objects (Fig.~\ref{fig:h0}), and 
accurately defines the slope in the Hubble diagram\index{Hubble diagram} from $0 < z < 0.1$ to 1\%. To measure $H_0$, \SNt{Ia} must still
be externally calibrated with Cepheids, and this calibration is the major limitation to measuring $H_0$ with \SNt{Ia}.
Two separate teams have analyzed the Cepheids and \SNt{Ia} but have obtained divergent values for the Hubble constant\index{Hubble constant}.  Saha \etal\ \cite{Saha2001} find $H_0 = 59 \pm 6 $, whereas Freedman \etal\ \cite{Freedman2001} find $H_0 = 71\pm 2 \pm (6 \rm{~systematic}) $.
Of the 12 \SNt{Ia} for which there are Cepheid distances\index{galaxy!distance!Cepheid} to the host galaxy (\SN{1895B}$^*$, \SN{1937C}$^*$, \SN{1960F}$^*$, \SN{1972E},
\SN{1974G}$^*$, \SN{1981B}, \SN{1989B}, \SN{1990N}, \SN{1991T}, \SN{1998eq}, \SN{1998bu}, and \SN{1999by}), four were observed by non-digital means (marked by
$^*$) and are best excluded from analysis on the grounds that non-digital photometry routinely has systematic errors
far greater than 0.1 mag. Jha \cite{Jha02} has compared the \SNt{Ia} distances using an updated version
of \MLCs\ to the Cepheid host galaxy distances measured by the two \HST\ teams. Using only the digitally observed \SNt{Ia}, he finds, using distances from the \SNt{Ia} project of Saha \etal\ \cite{Saha2001}, $H_0 = 66 \pm 3 \pm (7 \rm{~systematic})$ \kmsMpc. Applying the same analysis to the \Key\ distances by  
Freedman \etal\ \cite{Freedman2001}  gives $H_0 = 76 \pm 3 \pm (8 \rm{~systematic})$ \kmsMpc. This difference is not due to \SNt{Ia} errors, but rather to the
different ways the two teams have measured Cepheid distances with \H. The two values do overlap when the
systematic uncertainties are included, but it is still uncomfortable that the discrepancies are so large, particularly when some systematic uncertainties are common between the two teams.

At present, SNe provide the most convincing constraints\index{Hubble constant} with $H_0 \sim 70 \pm 10$ \kmsMpc. However, future work
on measuring $H_0$ lies not with the SNe but with the Cepheid calibrators, or possibly in using other primary distance
indicators such as \EPm\ or the Sunyaev-Zeldovich effect\index{Sunyaev-Zeldovich effect}.

\section{The Measurement of Acceleration}
The intrinsic brightness of \SNt{Ia} allow them to be discovered to $z > 1.5$
with current instrumentation
(while a comparably deep search for \type{II} SNe would only reach redshifts of 
$z \sim 0.5$).
In the 1980s, however, finding, identifying, and studying 
even the impressively luminous \type{Ia} SNe was a daunting challenge, even 
towards the lower end of the redshift range shown in Fig.~\ref{fig:DLvsz}.
At these redshifts, beyond $z \sim 0.25$, Fig.~\ref{fig:DLvsz} shows that 
relevant cosmological models could be distinguished by differences of order 0.2 mag 
in their predicted luminosity distances\index{cosmology!luminosity distance}. 
For \SNt{Ia} with a dispersion of 0.2 mag, 10 well 
observed objects should 
provide a 3$\sigma$ separation between the various cosmological models. It should be noted that the uncertainty described
above in measuring $H_0$\index{Hubble constant} is not important in measuring the 
parameters for different cosmological models.  Only the relative
brightness of objects near and far is being exploited in Eq.~(\ref{eq:DL}) and the absolute value of $H_0$ 
scales out.

The first distant SN search was started by the Danish team
of N{\o}rgaard-Nielsen \etal\ \cite{NN89}.  With significant effort and large amounts 
of telescope time spread over more than two years, they discovered a 
single \SNt{Ia} in a $z = 0.3$ cluster of galaxies (and one \SNt{II} at $z = 
0.2$) \cite{Hansen89,NN89}.  The \SNt{Ia} was discovered 
well after maximum light on an observing night that could not have been predicted, and was only marginally 
useful for cosmology\index{cosmology}.  However, it showed that such high redshift\index{supernova!high redshift} 
SNe did exist and could be found, but that they would be very difficult to 
use as cosmological tools.

Just before this first discovery in 1988, a search for high redshift 
\type{Ia} SNe using a then novel wide field camera on a much 
larger (4m) telescope was begun at the \LBNL\ and the \CPA, at 
Berkeley.  This search, now known as the \SCP, was inspired by the impressive studies of the late 1980s 
indicating that extremely similar \type{Ia} SN events could be 
recognized by their spectra and light curves, and by the success of the 
\LBNl\ fully robotic low-redshift SN search in finding 20 
SNe with automatic image analysis \cite{muller-SP,robotic92}.  

The \SCp\ targeted a much higher redshift range, $z > 0.3$, in order 
to measure the (presumed) deceleration of the Universe, so it faced a 
different challenge than the \CTSs\ search.  The high redshift\index{supernova!high redshift} 
SNe required discovery, spectroscopic confirmation, and 
photometric follow up on much larger telescopes.  This precious telescope time could neither be borrowed 
from other visiting observers and staff nor applied for in sufficient 
quantities spread throughout the year to cover all SNe 
discovered in a given search field, and with observations early 
enough to establish their peak brightness.  Moreover, since the 
observing time to confirm high redshift SNe was significant on 
the largest telescopes, there was a clear ``chicken and egg'' problem:  
telescope time assignment committees would not award follow-up time for 
a SN discovery that might, or might not, happen on a given run 
(and might, or might not, be well past maximum) and, without the follow-up 
time, it was impossible to demonstrate that high redshift SNe 
were being discovered by the \SCp.

By 1994, the \SCp\ had solved this problem, first by providing convincing 
evidence that SNe, such as \SN{1992bi}, could be discovered near 
maximum (and K-corrected) out to $z = 0.45$ \cite{SNat458},
and then by developing and successfully demonstrating a new observing 
strategy that could effectively  guarantee SN discoveries 
on a predetermined date, all before or near maximum light \cite{piau1,piau2,piau3,pthermo}.   Instead of discovering a single SN at a time 
on average (with some runs not finding one at all), the new 
approach aimed to discover an entire ``batch'' of half-a-dozen or more 
\type{Ia} SNe at a time by observing a much larger number of 
galaxies in a single two or three day period a few nights before new 
Moon.   By comparing these observations with the same observations 
taken towards the end of dark time almost three weeks earlier, it was 
possible to select just those SNe that were still on the rise or 
near maximum.   The chicken and egg problem was solved, and now the 
follow-up spectroscopy and photometry could be applied for and scheduled on a pre-specified set of nights.  The new strategy 
worked -- the \SCp\ discovered batches of high redshift SNe,and no one 
would ever again have to hunt for high-redshift SNe without
the crucial follow-up scheduled in advance.

The \HZSNS\ was conceived at the end of 1994, 
when this group of astronomers became convinced that it was both possible to discover \SNt{Ia} in large numbers at $z>0.3$ by the efforts 
of Perlmutter \etal \cite{piau1,piau2,piau3}, and also use them as precision 
distance indicators as demonstrated by the \CTSs\ group \cite{H95}. Since 1995, the \SCp\ and \HZSNs\ have both
worked avidly to obtain a significant set of high redshift \SNt{Ia}.  

\subsection{Discovering \SNt{Ia}}

The two high redshift teams both used this pre-scheduled 
discovery and follow-up batch strategy.  They each aimed to use the observing 
resources they had available to best scientific advantage, choosing, 
for example, somewhat different exposure times or filters.

Quantitatively, \type{Ia} SNe\index{supernova!rate} are rare events on an astronomer's 
time scale -- they occur in a 
galaxy like the \GAL{Milky Way} a few times per millennium (see, \eg \cite{Cappellaro97,pain96-SP,pain02} and the chapter by Cappellaro in this volume). With modern 
instruments on 4 meter-class telescopes, which observe 1/3 of a square 
degree to 
${\rm R} = 24$ mag in less than 10 minutes, it is possible to search a 
million 
galaxies to $z < 0.5$ for \SNt{Ia} in a single night. 

Since \SNt{Ia} take approximately 20 days to rise from undetectable to 
maximum light \cite{Riessrisetime}, the three-week separation 
between observing periods (which equates 
to 14 rest frame days at $z = 0.5$) is a good filter to catch the 
SNe on the rise.  The SNe are not always 
easily identified as new stars on the bright background of their host galaxies, so a relatively sophisticated process must be used to 
identify 
them.  The process, which involves 20 Gigabytes of imaging data per night, consists of aligning a previous epoch, matching the image 
star profiles
(through convolution), and scaling the two epochs to make 
the two images as 
identical as possible. The difference between these two images is then 
searched for new objects 
which stand out against the static sources that have been largely 
removed in 
the differencing process \cite{SNat458,Perl97,pthermo,S98}.   The dramatic increase in computing power 
in the 1980s was an important element in the development of 
this search technique, as was the construction of wide-field cameras 
with ever larger CCD detectors or mosaics of such detectors \cite{Wittman98}.

This technique is very efficient at producing large numbers of objects
that are, on average, at or near maximum light, and does not require unrealistic amounts of large telescope
time. It does, however, place the burden of work on follow-up observations, usually with different
instruments on different telescopes. With the large number of objects discovered (50 in 
two nights being typical), a new strategy is being adopted by both the \SCp\ and \HZSNs\ teams, as well as additional teams like the \CFHT\
legacy survey, where the same fields are repeatedly scanned several times per month, in multiple colors,
for several consecutive months. This type of observing program provides both discovery of new objects
and their follow up, all integrated into one efficient program. It does require a
large block of time on a single telescope -- a requirement which was not politically feasible
in years past, but is now possible.
 
\subsection{Obstacles to Measuring Luminosity Distances at High-$Z$}

As shown above, the distances measured to \SNt{Ia} are well characterized at $z < 0.1$,
but comparing these objects to their more distant counterparts requires
great care. Selection effects can introduce systematic errors as a function
of redshift, as can uncertain K-corrections and a possible evolution of the \SNt{Ia} progenitor
population as a function of look-back time. These effects, if they are large and not constrained
or corrected, will limit our ability to accurately measure relative luminosity 
distances\index{cosmology!luminosity distance}, and have the potential to reduce the efficacy of high-$z$ \type{Ia} SNe for measuring cosmology\index{cosmology} \cite{Perl97,Perl99,Riess98,S98}.

\subsubsection{K-Corrections\index{cosmology!K-correction}:}

As SNe are observed at larger and larger redshifts, their light is shifted to 
longer wavelengths. Since astronomical observations are normally made in fixed 
band passes on Earth, corrections need to be applied to account for the differences 
caused by the spectrum shifting within these band passes. These 
corrections take the form of integrating the spectrum of an SN over the relevant band passes, shifting the SN spectrum to the correct redshift,
and re-integrating. Kim \etal\ \cite{Kim96} 
showed that these effects can be minimized if one does not 
use a single bandpass, but instead chooses the bandpass closest to 
the redshifted rest-frame bandpass, as they had done for \SN{1992bi}
\cite{SNat458}. They showed that the inter-band K-correction is given by

\begin{equation}K_{ij}(z) = 2.5\log\left[(1+z) {\int F(\lambda) S_i(\lambda)d\lambda \over
\int F(\lambda/(1+z))S_j(\lambda)d\lambda)} {\int{Z}(\lambda) 
S_j(\lambda)d\lambda \over \int Z(\lambda)S_i(\lambda)d\lambda} \right], \label{eq:NewKcorr}
\end{equation}

\noindent where $K_{ij}(z)$ is the
correction to go from filter $i$ to filter $j$, and $Z(\lambda)$ is
the spectrum corresponding to zero magnitude of the filters. 

The brightness of an object expressed in magnitudes, as a function of $z$ is
\begin{equation}
m_i (z) = 5\log\left[{D_L(z) \over {\rm Mpc}}\right] + 25 +  M_j + K_{ij}(z), \label{eq:Kcorrdist} 
\end{equation}

\noindent where $D_L(z)$ is given by Eq.~(\ref{eq:DL}), $M_j$ is the absolute magnitude of object
in filter $j$, and $K_{ij}$ is given by Eq.~(\ref{eq:NewKcorr}). For example, for $H_0 = 70$ \kmsMpc,
and $D_L = 2835$ Mpc ($\Omega_M = 0.3, \Omega_\Lambda = 0.7$), at maximum light a \SNt{Ia} has $M_B = -19.5$ mag
and a $K_{BR} = -0.7$ mag.  We therefore expect an \SNt{Ia} at $z = 0.5$ to peak at $m_{\rm R} \sim 22.1$ mag for this set of cosmological
parameters.

K-correction errors depend critically on three uncertainties:

\begin{enumerate}
\item{Accuracy of spectrophotometry of SNe. To calculate the K-correction, the spectra of SNe
are integrated in Eq.~(\ref{eq:NewKcorr}). These integrals are insensitive to a grey shift
in the flux calibration of the spectra, but any wavelength dependent flux calibration error will translate
into erroneous K-corrections.}
\item{Accuracy of the absolute calibration of the fundamental astronomical standard 
systems. Eq.~(\ref{eq:NewKcorr}) shows that the K-corrections are sensitive to the shape
of the astronomical band passes and to the zero points of these band passes.} 
\item{Accuracy of the choice of \SNt{Ia} spectrophotometry template used to calculate the corrections. Although a relatively homogenous
class, there are variations in the spectra of \SNt{Ia}. If a particular object has, for example, a stronger
calcium triplet than the average \SNt{Ia}, the K-corrections will be in error unless an appropriate subset of \SNt{Ia}
spectra are used in the calculations.}
\end{enumerate}

The first error should not be an issue if correct observational procedures
are used on an instrument that has no fundamental problems. 
The second error is currently estimated to be small ($\sim0.01$ mag), 
based on the 
consistency of spectrophotometry and broadband photometry of the fundamental 
standards, Sirius\index{star!individual!Sirius} and Vega\index{star!individual!Vega} \cite{Bessell98}.  To
improve this uncertainty will require new, careful experiments to accurately calibrate a star, such
as Vega or Sirius (or a White Dwarf\index{star!white dwarf} or solar analog star\index{star!solar analog}), 
and to carefully infer the standard bandpass that defines the
photometric system in use at telescopes.  The third error requires a large 
database to match as closely as possible an SN with the 
spectrophotometry used to calculate the K-corrections. Nugent \etal\ \cite{Nugent02}
have shown that extinction and color are related and, by correcting the spectra
to force them to match the photometry of the SN needing K-corrections, that it is possible
to largely eliminate errors 1 and 3, even when using spectra that are not
exact matches (in epoch or in fine detail) to the \SNt{Ia} being K-corrected. Scatter in the measured 
K-corrections from a variety of telescopes and objects allows us to estimate the 
combined size of the effect for the first and third errors.  These appear to be $\sim0.01$ mag for redshifts where the high-$z$ and low-$z$ filters have a large region of overlap (\eg R-band matched to B-band at $z = 0.5$). 

\subsubsection{Extinction:}

In the nearby Universe we see \SNt{Ia} in a variety of environments, and about 10\% 
have significant extinction \cite{HP99}.
Since we can correct for extinction by observing two or more wavelengths, it is possible
to remove any first order effects caused by a changing average extinction of
\SNt{Ia} as a function of $z$. However, second order effects, such as possible 
evolution of the average properties of intervening dust, could still introduce
systematic errors. This problem can also be addressed by observing distant \SNt{Ia} over a decade or so of wavelength in order to measure the extinction law
to individual objects.  Unfortunately, this is observationally very expensive. Current observations 
limit the total systematic effect to $<0.06$ mag, as most of our current 
data is based on two color observations. 

An additional problem is the existence of a thin veil of dust around the \GAL{Milky Way}.  Measurements from the \COBE\ satellite accurately determined the relative
amount of dust around the Galaxy \cite{SFD98},
but there is an uncertainty in the absolute amount of extinction of about $2-3$\%.
This uncertainty is not normally a problem, since it affects 
everything in the sky more or less equally. However, as we observe SNe at higher 
and higher redshifts, the light from the objects is shifted to the red and is 
less affected by the Galactic dust. Our present knowledge indicates that a systematic error as large as 0.06 
mag is attributable to this uncertainty.
\subsubsection{Selection Effects:}

As we discover SNe, we are subject to a variety of selection effects, both in our 
nearby and distant searches. The most significant effect is the Malmquist Bias\index{Malmquist Bias} -- a 
selection effect which leads magnitude limited searches to find brighter than 
average objects near their distance limit since brighter objects can be seen in a 
larger volume than their fainter counterparts. Malmquist Bias errors are 
proportional to the square of the intrinsic dispersion of the distance method, 
and because \SNt{Ia} are such accurate distance indicators these errors are quite 
small, $\sim 0.04$ mag. Monte Carlo simulations can be used to estimate such selection effects, and to remove them from our data sets \cite{Perl97,pthermo,Perl99,S98}. The total uncertainty from selection effects
is $\sim0.01$ mag and, interestingly, may be worse for lower redshift objects because they are, at present,
more poorly quantified.

\subsubsection{Gravitational Lensing\index{cosmology!gravitational lensing}:}

Several authors have pointed out that the radiation from any object, as it 
traverses the large scale structure between where it was emitted and where it 
is detected, will be weakly lensed as it encounters fluctuations in the 
gravitational potential \cite{HW98,KVB95,Wam97}.
On average, most of the light travel paths go through 
under-dense regions and objects appear de-magnified. Occasionally, the light path encounters dense regions and the object becomes magnified.  The distribution of 
observed fluxes for sources is skewed by this process such that the vast 
majority of objects appear slightly fainter than the canonical luminosity 
distance, with the few highly magnified events making the mean of all light paths 
unbiased.  Unfortunately, since we do not observe enough objects to capture the 
entire distribution, unless we know and include the skewed shape of the 
lensing a bias will occur. At $z = 0.5$, this lensing is not a significant 
problem:  If the Universe is flat in normal matter, the large scale structure 
can induce a shift of the mode of the distribution by only a few percent. However, 
the effect scales roughly as $z^2$, and by $z = 1.5$ the effect can be as large as 
25\% \cite{Holz98}. While corrections can be derived by measuring
the distortion of background galaxies near the line of sight to each SN, at $z > 1$,
this problem may be one which ultimately limits the accuracy of luminosity distance 
measurements, unless a large enough sample of SNe at each redshift can be used to characterize the lensing distribution and average out the effect. For the $z \sim 0.5$ sample, the error is
$<0.02$ mag, but it is much more significant at $z > 1$ (\eg for \SN{1997ff})  \cite{Benitez2002,mortsell2001}, especially if the sample size is small.

\subsubsection{Evolution\index{cosmology!supernova evolution}:}

\SNt{Ia} are seen to evolve in the nearby Universe. Hamuy \etal\ \cite{H96a}
plotted the shape of the SN light curves against 
the type of host galaxy. SNe in early hosts 
(galaxies without recent star formation) consistently show light curves which 
rise and fade more quickly than SNe in late-type\index{galaxy!late type} hosts (galaxies with on-going star formation). However, once corrected for 
light curve shape the luminosity shows no bias as a function of 
host type.  This empirical investigation provides reassurance for using \SNt{Ia} as distance indicators over a variety of stellar population ages.  It is possible, of course, to devise
scenarios where some of the more distant SNe do not have nearby 
analogues, so as supernovae are studied at increasingly higher redshifts it can become important to obtain 
detailed spectroscopic and photometric observations of every distant SN to 
recognize and reject examples that have no nearby analogues.

In principle, it should be possible to use differences in the 
spectra and light curves between nearby and distant SNe, combined with theoretical modeling,  to correct
any differences in absolute magnitude.  Unfortunately,
theoretical investigations are not yet advanced enough to precisely quantify
the effect of these differences on the absolute magnitude. A different, empirical 
approach to handle SN evolution \cite{coping} is to divide the SNe into subsamples
of very closely matched events, based on the details of the their light curves, spectral
time series, host galaxy properties, etc.  A separate Hubble diagram\index{Hubble diagram} can then be constructed
for each subsample of SNe, and each will yield an independent measurement of the 
cosmological parameters.  The agreement (or disagreement) between the results from the
separate subsamples is an indicator of the total effect of evolution.  A simple, first attempt
at this kind of test has been performed by comparing the results
for SNe found in elliptical\index{galaxy!elliptical} host galaxies to SNe found in late spirals\index{galaxy!spiral} or
irregular hosts, and  the cosmological results from these subsamples were found to agree
well \cite{sullivan02}. 

Finally, it is possible to move to higher redshifts and see if the SNe deviate from the predictions
of Eq.~(\ref{eq:DL}). At a gross level, we expect an accelerating Universe\index{cosmology!Universe!acceleration} to be decelerating\index{cosmology!Universe!deceleration}
in the past because the matter density\index{cosmology!Universe!matter density} of the Universe increases with redshift, whereas the density of any
dark energy\index{cosmology!Universe!dark energy} leading to acceleration will increase at a slower rate than this (or not at all in the case of
a cosmological constant\index{cosmology!cosmological constant}). If the observed acceleration is caused by some sort of systematic effect, it is
likely to continue to increase (or at least remain steady) with $z$, rather than
disappear like the effects of dark energy. A first comparison has been made with \SN{1997ff} at $z\sim1.7$ \cite{Riess01}, and it seems consistent with a decelerating Universe at that epoch. More objects are necessary
for a definitive answer, and these should be provided by several large programs
that have been discovering such \type{Ia} SNe at the \KI, \SUBARU, and \H\ telescopes.

\subsection{High Redshift \SNt{Ia} Observations}

\begin{figure}[p]
\centering \includegraphics[width=12cm]{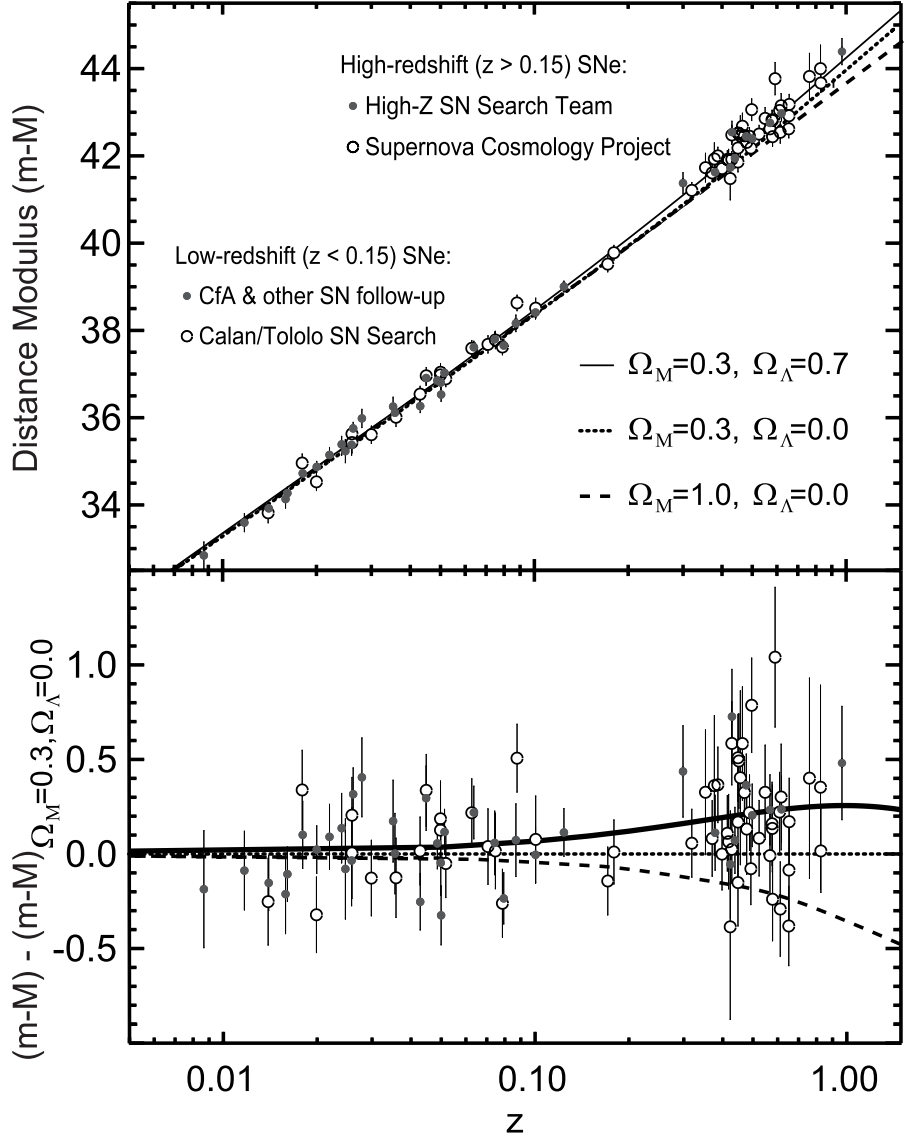}
\caption{{\em Upper panel:} The Hubble diagram for high redshift \SNt{Ia} from both the \HZSNs\ \cite{Riess98} and the \SCp\ 
\cite{Perl99}.  {\em Lower panel:} The residual of the distances relative to 
a $\Omega_M = 0.3$, $\Omega_\Lambda = 0.7$ Universe. The $z<0.15$ objects for both 
teams are drawn from \CTSs\ sample \cite{H95}, so many of 
these objects are in common between the analyses of the two teams.}
\label{fig:hubdiagramhighz}
\end{figure}

\begin{figure}[p]
\centering \includegraphics[width=12cm]{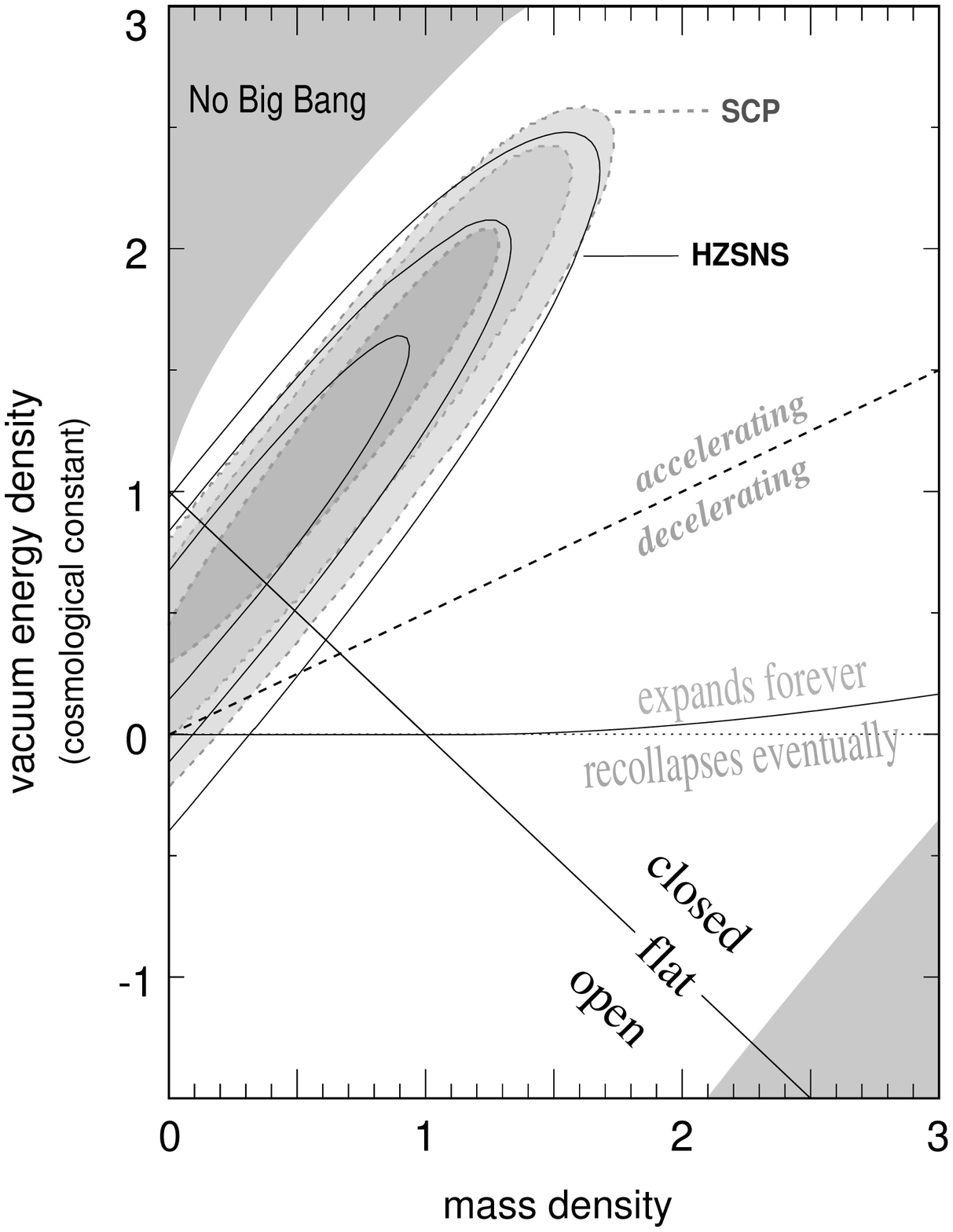}
\caption{The confidence regions for both \HZSNs\ \cite{Riess98} and \SCp\ \cite{Perl99} for
$\Omega_M$, $\Omega_\Lambda$. The two experiments show, with remarkable consistency, that $\Omega_\Lambda > 0$ is required to
reconcile observations and theory. 
The \SCp\ result is based on measurements of 42 distant \SNt{Ia}.  (The analysis shown here is uncorrected for host galaxy extinction;see \cite{Perl99} for the alternative 
analyses with host extinction correction, which is shown to make little 
difference in this data set.)  The \HZSNs\ result is based on measurements of
16 \SNt{Ia}, including 6 snapshot distances \cite{Riesssnapshot}, of which 
two are \SCp\ SNe from the 42 SN sample.
The $z < 0.15$ objects used to constrain the fit for both teams are drawn from 
the \CTSs\ sample \cite{H95},
so many of these objects are common between the analyses by the two teams.}
\label{fig:highzcont}
\end{figure}

The \SCp\ \cite{Perl97} in 1997 presented their first results with 7 objects at a redshift 
around $z = 0.4$. These objects hinted at a decelerating 
Universe with a measurement of $\Omega_M = 0.88 ^{+0.69}_{-0.60}$,
but were not definitive. Soon after, the \SCp\ published a further result, 
with a $z \sim 0.84$ \SNt{Ia} observed with 
the \Ki\ and \H\ added to the sample \cite{Perl98}, and
the \HZSNs\ presented the results from their first four objects \cite{Garn98a,S98}.  The results from both teams now ruled out a $\Omega_M = 1$ Universe
with greater than 95\% significance. These findings were again superceded 
dramatically when both teams announced results including more SNe (10 more \HZSNs\ SNe, and 34 more \SCp\ SNe) that showed not
only were the SN observations incompatible with a $\Omega_M = 1$ Universe, they were also incompatible
with a Universe containing only normal matter \cite{Perl99,Riess98}.
Fig.~\ref{fig:hubdiagramhighz} shows the Hubble diagram\index{Hubble diagram} for both teams. Both samples show that SNe 
are, on average, fainter than would be expected, even for an empty 
Universe, indicating that the Universe is accelerating\index{cosmology!Universe!acceleration}. The agreement between the experimental results of the two teams is spectacular, especially considering the two programs have worked in almost
complete isolation from each other.

The easiest solution to explain the observed acceleration is to include an additional 
component of matter with an equation of state parameter more negative than 
$w<-1/3$; the most familiar being the cosmological constant\index{cosmology!cosmological constant} ($w = -1$).  Fig.~\ref{fig:highzcont}
shows the joint confidence contours for values of $\Omega_M$ and $\Omega_\Lambda$ from both 
experiments. If we assume the Universe is composed only of normal matter and a cosmological constant, then with 
greater than 99.9\% confidence the Universe has a non-zero cosmological constant or some other form of dark energy\index{cosmology!Universe!dark energy}.

\begin{figure}[t]
\centering
\centering \includegraphics[width=12cm]{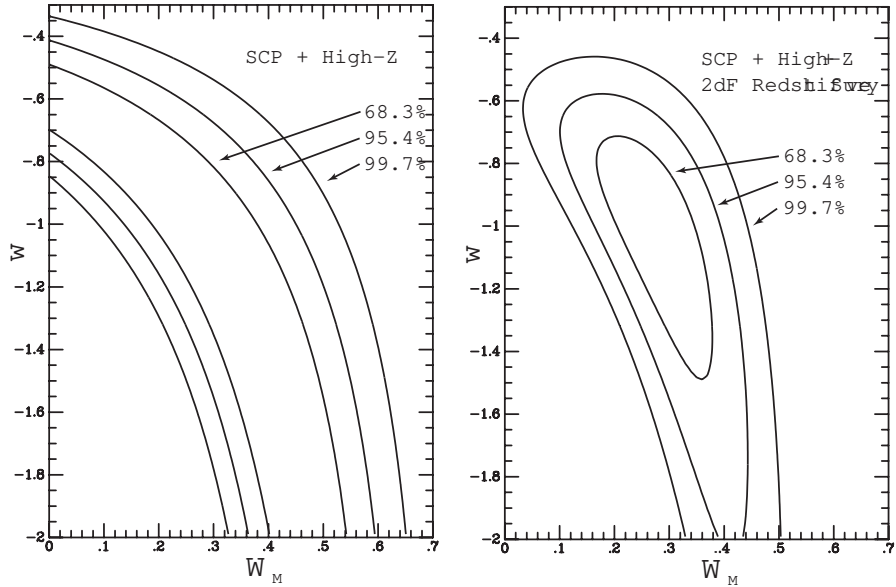}
\caption{Left panel: Contours of $\Omega_M$ versus $w_x$ from current observational data.  Right Panel: Contours of $\Omega_M$ versus $w_x$ from current observational data, where the current value of $\Omega_M$ is obtained from the 2dF redshift survey.  For both panels $\Omega_M + \Omega_x = 1$ is taken as a prior.}
\label{fig:currentw}
\end{figure}

\begin{figure}[ht]
\centering
\centering \includegraphics[width=12cm]{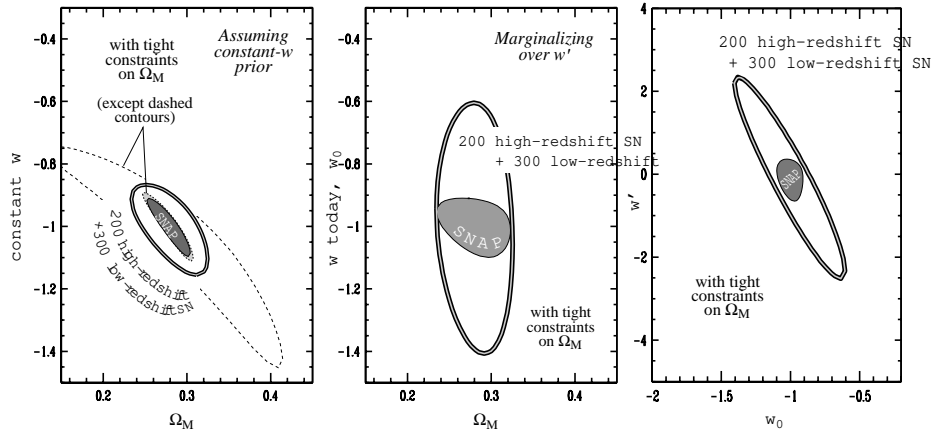}
\caption{Future expected constraints on dark energy: {\em Left panel:} Estimated 68\% confidence regions for a 
constant equation of state parameter for 
the dark energy, $w$, versus mass density, for a ground-based 
study with 200 SNe between $z = 0.3-0.7$ (open contours), and for the 
satellite-based \SNAp\ experiment with 2,000 SNe between $z = 0.3-1.7$ 
(filled contours).  Both experiments are assumed to also use 300 
SNe between $z = 0.02-0.08$.
A flat cosmology\index{cosmology} is assumed (based on \CMB\ constraints) and the inner  
(solid line) contours for each experiment 
include tight constraints (from large scale structure surveys) 
on $\Omega_M$, at the $\pm 0.03$ level.  For the \SNAp\
experiment, systematic uncertainty is taken as $dm = 0.02 (z/1.7)$, and for 
the ground-based experiment, $dm = 0.03 (z/0.5)$.  Such ground-based studies
will test the hypothesis that the dark energy is in the form of a
cosmological constant, for which $w = -1$ at all times. {\em Middle panel:}  The same confidence regions for the same experiments
not assuming the equation of state parameter, $w$, to be 
constant, but instead marginalizing over $w^\prime$, where $w(z) = w_0 +
w^\prime z$. (Weller and Albrecht
\cite{welleralbrecht} recommend this parameterization 
of $w(z)$ over the others that have been proposed to characterize well the 
current range of dark energy models.)  Note that these planned ground-based 
studies will yield impressive constraints on the value of $w$ today, $w_0$,
even without assuming constant $w$.  In fact, these constraints are comparable 
to the current measurements of $w$ assuming it is constant (shown 
in the right panel of Fig. 6). {\em Right panel:}  Estimated 68\% confidence regions of the first 
derivative of the equation of state, $w^\prime$, versus its value today, $w_0$, 
for the same experiments. 
} \label{fig:future}
\end{figure}

Of course, we do not know the form of dark energy\index{cosmology!Universe!dark energy} which is leading to the 
acceleration\index{cosmology!Universe!acceleration}, and it is worthwhile investigating what other forms of energy are
possible additional components. Fig.~\ref{fig:currentw} shows the joint confidence 
contours for the \HZSNs+\SCp\ observations for $\Omega_M$ and $w_x$ (the equation of state 
of the unknown component causing the acceleration).  Because this 
introduces an extra parameter, we apply the additional constraint that 
$\Omega_M + \Omega_x = 1$, as indicated by the \CMb\
experiments \cite{deBern00}. The cosmological constant\index{cosmology!cosmological constant} is preferred, but anything with a 
$w < -0.5$ is acceptable \cite{Garn98b,Perl99}.
Additionally, we can add information about the value of $\Omega_M$, 
as supplied by recent 2dF redshift survey results \cite{verde02}, as shown in the 2nd panel,
where the constraint strengthens to $w<-0.6$ at 95\% confidence \cite{PTW01}.

\section{The Future}

How far can we push the SN measurements? Finding more and more SNe allows us to 
beat down statistical errors to arbitrarily small levels but, ultimately, systematic effects will limit the precision to which \SNt{Ia} magnitudes can 
be applied to measure distances. Our best estimate is that it will be possible to
control systematic effects\index{cosmology!systematic effects} from ground-based experiments to a level of $\sim0.03$ mag.
Carefully controlled ground-based experiments on 200 SNe will reach this statistical
uncertainty in $z = 0.1$ redshift bins in the range $z = 0.3-0.7$, and is achievable within five years.  A comparable quality
low redshift sample, with 300 SNe in $z = 0.02-0.08$, will also be achievable
in that time frame \cite{SNfactory}.

The \SNAP\ collaboration\footnote{See \emph{http://snap.lbl.gov}} has proposed 
to launch a dedicated cosmology\index{cosmology}
satellite \cite{SNAPaldering,SNAPperl} -- the ultimate \SNt{Ia} experiment. This satellite will, if funded, scan many square 
degrees of sky, discovering well over a thousand \SNt{Ia} per year and obtain their spectra and 
light curves out to $z = 1.7$.  Besides the large numbers of objects
and their extended redshift range, space-based observations will also provide the opportunity to control
many systematic effects better than from the ground \cite{frieman02,lindhuter}. Fig.~\ref{fig:future} shows the expected
precision in the \SNAp\ and ground-based experiments for measuring $w$, assuming
a flat Universe\index{cosmology!Universe!curvature}.  Perhaps the most important advance will be the first studies 
of the time variation of the equation of state $w$ (see the right panel of 
Fig.~\ref{fig:future} and \cite{hutturn01,welleralbrecht}).

With rapidly improving \CMb\ data from interferometers, the satellites \MAP\ and 
\PLANCK, and balloon-based instrumentation planned for the next several years, 
\CMb\ measurements promise dramatic improvements in precision on many of the
cosmological parameters.  However, the \CMb\ measurements are relatively 
insensitive to the dark energy and the epoch of cosmic acceleration.  
\SNt{Ia} are currently the only way to directly study this acceleration epoch with sufficient
precision (and control on systematic uncertainties) that we can investigate the
properties of the dark energy, and any time dependence in these properties.
This ambitious goal will require complementary and supporting measurements
of, for example, $\Omega_M$ from \CMb,  weak lensing, and large scale structure. 
The SN measurements will also provide a test of the cosmological results 
independent from these other techniques, which have their own systematic errors.  Moving forward
simultaneously on these experimental fronts offers the plausible and exciting 
possibility of achieving a comprehensive measurement of the fundamental 
properties of our Universe.




\begin{thebibliography}{22.}
\addcontentsline{toc}{section}{References}

\bibitem{SNAPaldering} G.~Aldering \etal: \spie \textbf{4835} 21 (2002)
\bibitem{SNfactory} G.~Aldering \etal: \spie \textbf{4836} 93 (2002)
\bibitem{B57} W.A.~Baum:  \aj  \textbf{62} 6 (1957)
\bibitem{Benitez2002} N.~Benitez, A.~Riess, P.~Nugent, M.~Dickinson, R.~Chornock, A.~Filippenko:  \apjl \textbf{577} L1 (2002)
\bibitem{Bessell98} M.~Bessell: \pasp \textbf{102} 1181 (1998)
\bibitem{branch89} D.~Branch: In: \emph{Encyclopedia of Astronomy and Astrophysics} (Academic, San Diego 1989) p.~733
\bibitem{branchtammann92} D.~Branch, G.A.~Tammann: \araap \textbf{30} 359 (1992)
\bibitem{branchfisher93} D.~Branch, A.~Fisher, P.~Nugent: \aj \textbf{106} 2383 (1993)
\bibitem{B96} D.~Branch, A.~Fisher, E.~Baron, P.~Nugent: \apjl \textbf{470} L7 (1996)
\bibitem{coping} D.~Branch, S.~Perlmutter, E.~Baron, P.~Nugent: In: \emph{Resource Book on Dark Energy}, ed.~by E.V.~Linder (Snowmass 2001)
\bibitem{cadonau87} R.~Cadonau: PhD Thesis, University of Basel (1987)
\bibitem{Cappellaro97} E.~Cappellaro, M.~Turatto, D.Yu.~Tsvetkov, O.S.~Bartunov, C.~Pollas, R.~Evans, M.~Hamuy: \aap \textbf{322} 431 (1997)
\bibitem{CL} P.~Coles, F.~Lucchin: In: \emph{cosmology\index{cosmology}}  (Wiley, Chicester 1995) p.~31
\bibitem{deBern00} P.~de Bernardis \etal:  \nat \textbf{404} 955 (2000)
\bibitem{EK89} R.G.~Eastman, R.P.~Kirshner: \apj \textbf{347} 771 (1989)
\bibitem{ESK96} R.G.~Eastman, B.P.~Schmidt, R.~Kirshner: \apj \textbf{466} 911 (1996)
\bibitem{F95} A.~Fisher, D.~Branch, P.~Hoeflich, A.~Khokhlov: \apjl \textbf{447} L73 (1995)
\bibitem{filippenko91} A.V.~Filippenko: In: \emph{SN1987A and Other Supernovae}, ed.~by I.J.~Danziger, K.~Kjar (ESO, Garching 1991) p.~343
\bibitem{Fetal92} A.V.~Fillipenko \etal: \apjl \textbf{384} L15 (1992)
\bibitem{Freedman2001} W.L.~Freedman \etal: \apj \textbf{553} 47 (2001)
\bibitem{frieman02} J.~Frieman, D.~Huterer, E.V.~Linder, M.S.~Turner: \aph 0208100 (2002)
\bibitem{Garn98a} P.~Garnavich \etal: \apjl \textbf{493} L53 (1998)
\bibitem{Garn98b} P.~Garnavich \etal: \apj \textbf{509} 74 (1998)
\bibitem{Germany02} L.G.~Germany, A.G.~Riess, B.P.~Schmidt, N.B.~Suntzeff: \inprep (2003)
\bibitem{goldhaber97} G.~Goldhaber \etal: In: \emph{Thermonuclear Supernovae}, ed.~by P.~Ruiz-Lapuente, R.~Canal, J.~Isern (Aiguablava, June 1995; NATO ASI, 1997)
\bibitem{goldhaber01} G.~Goldhaber \etal: \apj \textbf{558} 359 (2001)
\bibitem{GoobPerl95} A.~Goobar, S.~Perlmutter: \apj  \textbf{450} 14 (1995)
\bibitem{hamuy91} M.~Hamuy, M.M.~Phillips, J.~Maza, M.~Wischnjewsky, A.~Uomoto, A.U.~Landolt, R.~Khatwani: \aj \textbf{102} 208 (1991)
\bibitem{H96a} M.~Hamuy, M.M.~Phillips, N.B.~Suntzeff, R.A.~Schommer, J.~Maza, R.~Aviles: \aj \textbf{112} 2391 (1996)
\bibitem{HP99} M.~Hamuy, P.A.~Pinto: \aj \textbf{117} 1185 (1999)
\bibitem{H93} M.~Hamuy \etal: \aj \textbf{106} 2392 (1993)
\bibitem{H95} M.~Hamuy \etal: \aj \textbf{109} 1 (1995)
\bibitem{H96c} M.~Hamuy \etal: \aj \textbf{112} 2408 (1996)
\bibitem{h2001} M.~Hamuy \etal: \apj \textbf{558} 615 (2001) 
\bibitem{Hansen89} L.~Hansen, H.E.~Jorgensen, H.U.~N{\o}rgaard-Nielsen, R.S.~Ellis, W.J.~Couch: \aap \textbf{211} L9 (1989)
\bibitem{harkness90} R.P.~Harkness, J.C.~Wheeler: In: \emph{Supernovae}, ed.~by A.G.~Petschek (Springer-Verlag, New York 1990) p.~1
\bibitem{HW98} D.E.~Holz, R.M.~Wald: \phyrd \textbf{58} 063501 (1998)
\bibitem{Holz98} D.E.~Holz: \apj \textbf{506} 1 (1998)
\bibitem{HMS56} M.L.~Humason, N.U.~Mayall, A.R.~Sandage: \apj \textbf{61} 97 (1956)
\bibitem{hutturn01} D.~Huterer, M.S.~Turner: \phyrd \textbf{64} 123527  (2001)
\bibitem{Jha02} S.~Jha: PhD Thesis, Harvard University (2002)
\bibitem{K68} C.T.~Kowal: \aj \textbf{73} 1021 1968
\bibitem{Kim96} A.~Kim, A.~Goobar, S.~Perlmutter: \pasp  \textbf{108} 190 (1996)
\bibitem{KK74} R.P.~Kirshner, J.~Kwan: \apj \textbf{193} 27 (1974)
\bibitem{KVB95} R.~Kantowski, T.~Vaughan, D.~Branch: \apj \textbf{447} 35 (1995)
\bibitem{leib88} B.~Leibundgut: PhD Thesis, University of Basel (1988) 
\bibitem{leibtammann90} B.~Leibundgut, G.A.~Tammann: \aap  \textbf{230} 81 (1990)
\bibitem{leibetal91} B.~Leibundgut, G.A.~Tammann, R.~Cadonau, D.~Cerrito: \aas  \textbf{89} 537 (1991)
\bibitem{Letal93} B.~Leibundgut \etal: \aj \textbf{105} 301 (1993)
\bibitem{Letal96} B.~Leibundgut \etal: \apjl \textbf{466} L21 (1996)
\bibitem{leonard2001} D.C.~Leonard, A.V.~Filippenko, D.R.~Ardila, M.S.~Brotherton: \apj \textbf{553} 861 (2001)
\bibitem{leonard2002} D.C.~Leonard \etal: \pasp \textbf{114} 35 (2002)
\bibitem{lindhuter} E.~Linder, D.~Huterer: \aph 0208138 (2002)
\bibitem{millerbranc90} D.L.~Miller, D.~Branch: \aj \textbf{100} 530 (1990)
\bibitem{mortsell2001} E.~Mortsell, C.~Gunnarsson, A.~Goobar: \apj \textbf{561} 106 (2001)
\bibitem{muller-SP} R.A.~Muller, H.J.M.~Newberg, C.R.~Pennypacker, S.~Perlmutter, T.P.~Sasseen, C.K.~Smith: \apjl \textbf{384} L9 (1992)
\bibitem{NN89} H.U.~N{\o}rgaard-Nielsen, L.~Hansen, H.E.~Jorgensen, A.~Aragon Salamanca, R.S.~Ellis: \nat \textbf{339} 523 (1989)
\bibitem{Nugent02} P.~Nugent, A.~Kim, S.~Perlmutter: \pasp \textbf{114} 803 (2002)
\bibitem{panagia85} N.~Panagia: In: \emph{Supernovae as Distance Indicators}, ed.~by N. Bartel, (Springer-Verlag, Berlin 1985) p.~14
\bibitem{pain96-SP} R.~Pain \etal: \apj \textbf{473} 356 (1996)
\bibitem{pain02} R.~Pain \etal: \apj \textbf{577} 120 (2002)
\bibitem{pearce88b} G.~Pearce, B.~Patchett, J.~Allington-Smith, I.~Parry: \apss \textbf{150} 267 (1988)
\bibitem{P87} M.M.~Phillips \etal: \pasp \textbf{99} 592 (1987)
\bibitem{P93} M.M.~Phillips: \apjl \textbf{413} L105 (1993)
\bibitem{Phillips99} M.M.~Phillips, P.~Lira, N.B.~Suntzeff, R.A.~Schommer, M.~Hamuy, J.~Maza: \aj \textbf{118} 1766 (1999)
\bibitem{Peebles93} P.J.E.~Peebles: In: \emph{Principles of Physical cosmology\index{cosmology}} (Princeton University Press, Princeton 1993)
\bibitem{robotic92} S.~Perlmutter, R.A.~Muller, H.J.M.~Newberg, C.R.~Pennypacker, T.P.~Sasseen, C.K.~Smith: \aspc \textbf{34} 67 (1992)
\bibitem{SNAPperl} S.~Perlmutter, E.~Linder: In \emph{Dark Matter 2002, Proc.~5th International UCLA Symposium on Sources and Detection of Dark Matter and Dark Energy in the Universe}, ed.~by D.B.~Cline (Elsevier, Amsterdam 2003)
\bibitem{PTW01} S.~Perlmutter, M.~Turner, M.~White: \prl \textbf{83} 670 (1999)
\bibitem{piau1} S.~Perlmutter \etal: \iauc 5956  (1994)
\bibitem{piau2} S.~Perlmutter \etal: \iauc 6263  (1995)
\bibitem{piau3} S.~Perlmutter \etal: \iauc 6270  (1995)
\bibitem{SNat458} S.~Perlmutter \etal: \apjl \textbf{440} L41 (1995)
\bibitem{Perl97} S.~Perlmutter \etal: \apj \textbf{483} 565 (1997)
\bibitem{Perl98} S.~Perlmutter \etal: \nat \textbf{391} 51 (1998)
\bibitem{pthermo} S.~Perlmutter \etal: In: \emph{Thermonuclear Supernovae}, ed.~by P.~Ruiz-Lapuente, R.~Canal, J.~Isern (Aiguablava, June 1995; NATO ASI, 1997)
\bibitem{Perl99} S.~Perlmutter \etal: \apj \textbf{517} 565 (1999)
\bibitem{Petal92} M.M.~Phillips, L.A.~Wells, N.B.~Suntzeff, M.~Hamuy, B.~Leibundgut, R.P.~Kirshner, C.B.~Foltz: \aj \textbf{103} 1632 (1992)
\bibitem{RPK96} A.G.~Riess, W.H.~Press, R.P.~Kirshner: \apj \textbf{473} 88 (1996)
\bibitem{Riesssnapshot} A.G.~Riess, P.~Nugent, A.V.~Filippenko, R.P.~Kirshner, S.~Perlmutter: \apj \textbf{504} 935 (1998)
\bibitem{Riessrisetime} A.G.~Riess, A.V.~Filippenko, W.~Li, B.P.~Schmidt: \aj \textbf{118} 2668 (1999) 
\bibitem{Riess97} A.G.~Riess \etal: \aj \textbf{114} 722 (1997)
\bibitem{Riess98} A.G.~Riess \etal: \aj \textbf{116} 1009 (1998)
\bibitem{Riess99a} A.G.~Riess \etal: \aj \textbf{117} 707 (1999)
\bibitem{Riess01} A.G.~Riess \etal: \apj \textbf{560} 49 (2001)
\bibitem{S94} B.P.~Schmidt, R.P.~Kirshner, R.G.~Eastman, M.M.~Phillips, N.B.~Suntzeff, N.B.~Hamuy, J.~Maza, R.~Aviles: \apj \textbf{432} 42 (1994)
\bibitem{S98} B.~Schmidt \etal: \apj \textbf{507} 46 (1998)
\bibitem{Saha2001} A.~Saha, A.~Sandage, G.A.~Tammann, A.E.~Dolphin, J.~Christensen, N.~Panagia, F.D.~Macchetto: \apj \textbf{562} 313 (2001) 
\bibitem{SFD98} D.J.~Schlegel, D.P.~Finkbeiner, M.~Davis: \apjs \textbf{500} 525 (1998)
\bibitem{ST93} A.~Sandage, G.A.~Tammann: \apj \textbf{415} 1 (1993)
\bibitem{sullivan02} M.~Sullivan \etal: \mnras { } {in press} (2003)
\bibitem{TL90} G.A.~Tammann, B.~Leibundgut: \aap \textbf{236} 9 (1990)
\bibitem{TS95} G.A.~Tammann, A.~Sandage: \apj \textbf{452} 16 (1995)
\bibitem{uomoto85} A.~Uomoto, R.P.~Kirshner: \aap \textbf{149} L7 (1985)
\bibitem{v95} S.~van den Bergh: \apjl \textbf{453} L55 (1995)
\bibitem{vBP92} S.~van den Bergh, J.~Pazder: \apj \textbf{390} 34 (1992)
\bibitem{vaughn95} T.E.~Vaughan, D.~Branch, D.L.~Miller, S.~Perlmutter: \apj \textbf{439} 558 (1995)
\bibitem{verde02} L.~Verde \etal: \mnras \textbf{335} 432 (2002)
\bibitem{wagoner81} R.V.~Wagoner: \apjl \textbf{250} L65 (1981)
\bibitem{Wam97} J.~Wambsgabss, R.~Cen, X.~Guohong, J.~Ostriker: \apjl \textbf{475} L81 (1997)
\bibitem{Wang2001} L.~Wang, A.D.~Howell, P.~H\"oflich, J.C.~Wheeler: \apj \textbf{550} 1030 (2001)
\bibitem{wheellev85} J.C.~Wheeler, R.~Levreault: \apjl \textbf{294} L17 (1985)
\bibitem{welleralbrecht} J.~Weller, A.~Albrecht: \phyrd \textbf{65} 103512 (2002)
\bibitem{Wittman98} D.M.~Wittman, J.A.~Tyson, G.M.~Bernstein, R.W.~Lee, I.P.~dell'Antonio, P.~Fischer, D.R.~Smith, M.M.~Blouke: \spie \textbf{3355} 626 (1998)
\end{thebibliography}
\end{document}